\documentclass{article}
\usepackage[latin1]{inputenc}
\usepackage[english]{babel}
\usepackage{dsfont}
\usepackage[cyr]{aeguill}
\usepackage{amsmath,amssymb}
\usepackage{amsthm}
\usepackage{eurosym}
\usepackage{multicol}
\usepackage{color}
\usepackage{fancyhdr}
\usepackage[pdftex]{graphicx}
\usepackage{bm} 
\usepackage[pdftex]{hyperref}
\usepackage{verbatim}
\usepackage{setspace}
\usepackage{stmaryrd}

\newcommand{\proba}{\mathbb{P}} 
\newcommand{\indic}{\mathds{1}} 
\newcommand{\E}{\mathbb{E}} 
\DeclareMathOperator{\diag}{diag} 

\theoremstyle{plain}
\newtheorem{thm}{Theorem}
\newtheorem{pro}{Proposition}

\newtheorem{lem}{Lemma}

\providecommand{\keywords}[1]{\textbf{\textit{Keywords --}} #1}

\title{Asymptotic and finite-sample distributions \\ of one- and two-sample empirical relative entropy}

\author{Matthieu Garcin$^{\text{a,}}$\thanks{Corresponding author: matthieu.garcin@m4x.org. \newline 
$^{\text{a}}$ De Vinci Higher Education, De Vinci Research Center, Paris, France. \newline
$^{\text{b}}$ Département de mathématiques et applications, \'Ecole normale sup\'erieure, 45 rue d'Ulm, 75005 Paris, France. \newline
Acknowledgements: MG acknowledges the support of the Chair ``Deep Finance Statistics'' between QRT, Ecole Polytechnique and its foundation. The authors would like to thank Olivier Benhamou for useful discussions and support.}, Louis Perot$^{\text{b}}$}

\date{\today}

\begin{document}

\maketitle

\begin{abstract}
In the perspective of building statistical tests of divergence between two probability distributions, we study the distribution of empirical relative entropy and derive several types of approximations: concentration inequalities for finite samples, asymptotic distributions, and Berry-Esseen bounds in a pre-asymptotic regime. For the latter, we introduce a new approach to obtain Berry-Esseen inequalities for nonlinear functions of sum statistics under some convexity assumptions. Our theoretical contributions cover both one- and two-sample empirical relative entropies. 
\end{abstract}

\keywords{Berry-Esseen bounds, concentration inequalities, information theory, Kullback-Leibler divergence, two-sample divergence testing}



\section{Introduction}

Various metrics can be used to build statistical tests of divergence between two probability distributions. One can cite for example Wasserstein distance, Hellinger distance, Kolmogorov-Smirnov statistic, or relative entropy~\cite{GS,GKL,Kelbert}. The latter metric is of particular interest since it is the expectation of the log-likelihood ratio. This ratio is the statistic leading to the uniformly highest power among the statistical tests of probability divergence, under the assumptions of Neyman-Pearson lemma~\cite{EC}.

Most of the divergence tests that use entropy consist in the comparison of an empirical distribution with a parametric one~\cite{Vasicek}. In some practical contexts, like change-point detection~\cite{LYC,HKB}, two empirical probabilities are to be compared. Beyond issues related to the consistency and bias of relative entropy estimators~\cite{WKV,BD}, the two-sample setting raises another difficulty. Characterizing the exact distribution of the empirical relative entropy is challenging and typically leads to statistical tests calibrated using thresholds based exclusively on simulated quantiles~\cite{PA}.

In this paper, we consider two discrete probability distributions, with a finite number of possible states. In the perspective of building a statistical test of equality of these distributions, we propose three types of approximations of the distribution of one- and two-sample empirical relative entropy.

We consider first the one-sample framework. The most natural approximation is the asymptotic distribution, which may however not be relevant when considering small samples, as it often appears in the context of change-point detection. We obtain as well pre-asymptotic and finite-sample bounds of the distribution, based either on a Berry-Esseen approach or on concentration inequalities. One could then use these bounds, instead of the asymptotic approximation, to build conservative statistical tests of divergence. Our Berry-Esseen bounds are obtained for a nonlinear function of a sum statistic, whose limit distribution is non-Gaussian. Our inequality controls two effects: the classical speed of convergence for an approximation of our statistic and the error related to this approximation. This method can easily be replicated to other kinds of nonlinear statistics. 

We propose as well extensions to two samples, that is approximations of the distribution of the relative entropy between two empirical probabilities, which is particularly useful in the context of change-point detection. This question is challenging because relative entropy does not satisfy a triangle inequality and because its empirical version may become unbounded when the reference probability is estimated from few observations.

The paper is organized as follows. Section~\ref{sec:empiricalKL} briefly introduces the statistic considered in this work, namely empirical relative entropy. Section~\ref{sec:CLT} deals with asymptotic and pre-asymptotic distributions, that is the central limit theorem and Berry-Esseen inequality in the context of empirical relative entropy. A comparison of finite-sample bounds, namely concentration inequalities, is presented in Section~\ref{sec:concentration}. Section~\ref{sec:conclusion} concludes.

\section{The statistic: empirical relative entropy}\label{sec:empiricalKL}

We consider a discrete probability, with a finite number $k\geq 2$ of possible categories: $\mathbf p=(p_1,...,p_k)^t\in(0,1)^k$, $Z^t$ standing for the transposed vector of $Z$. The Shannon entropy related to this categorical distribution is
$$H(\mathbf p)=-\sum_{i=1}^k p_i\log(p_i),$$
where we use the convention $0\log(0)=0$~\cite{CT}. The entropy quantifies the uncertainty of the distribution~\cite{CT,Garcin2024}. The minimum uncertainty corresponds to a concentration in a single state, leading to the minimum entropy, $H(\mathbf p)=0$. The maximum entropy is reached by a uniform distribution, for which we get $H(\mathbf p)=\log(k)$.

When working with data, we can calculate an empirical entropy, based on empirical probabilities. We observe $X_1$, ..., $X_n$, iid random variables, which may be either discrete or continuous. We discretize these variables by defining $k$ possible states $\Omega_1$, ..., $\Omega_k$, which may for example be intervals. We have $\proba(X_j\in\Omega_i)=p_i$ for all $j\in\llbracket 1,n\rrbracket$ and $i\in\llbracket 1,k\rrbracket$. We also define the empirical probability $\widehat{\mathbf p}_n=(\widehat p_{n,1},...,\widehat p_{n,k})^t\in[0,1]^k$, such that
\begin{equation}\label{eq:probaempirique}
\widehat p_{n,i}=\frac{1}{n}\sum_{j=1}^{n}\indic_{X_j\in\Omega_i}.
\end{equation}
The quantity $H(\widehat{\mathbf p}_n)$ is then the empirical entropy. The asymptotic distribution of the empirical entropy is either a chi-squared or a Gaussian distribution, depending on the nature of $\mathbf p$~\cite{Basharin,Zubkov}. 

One can also replace the probability $p$ by a conditional probability. It leads to the evaluation of the complexity of the dependence structure between two variables, what has been shown to be useful for time series, the presence of serial dependence being a useful asset for forecasting purposes~\cite{BG2024}. It has been shown that the distribution of conditional Shannon entropy and of the close concept of mutual information is similar to the one of the non-conditional entropy~\cite{LZ,SMM,BG2023,MB}.

Relative entropy, sometimes also called Kullback-Leibler divergence, uses the concept of entropy to compare to each other two probability distributions, $\mathbf p,\mathbf q\in(0,1)^k$:
$$D_{\text{KL}}\left(\mathbf p\Vert \mathbf q\right)=\sum_{i=1}^k p_i\log\frac{p_i}{q_i}.$$
Relative entropy is non-negative and not bounded. But, with a finite number of states, the infinity of the relative entropy is equivalent to the existence of a state $i$ for which $p_i\neq 0$ and $q_i=0$. Relative entropy is not symmetric in $\mathbf p$ and $\mathbf q$. When $q$ is uniform, relative entropy more simply writes $D_{\text{KL}}\left(\mathbf p\Vert \mathbf q\right)=\log(k)-H(\mathbf p)$. 

Again, replacing $\mathbf p$ by its empirical counterpart $\widehat{\mathbf p}_n$ leads to an empirical relative entropy. But one can also use relative entropy to compare to each other two empirical probabilities. We deal with the two-sample framework in this paper, considering that the two datasets are generated in the same distribution $\mathbf p$. We are thus given iid observations $X_1,...,X_{n+m}$ with $\proba(X_j\in\Omega_i)=p_i$ for all $j\in\llbracket 1,n+m\rrbracket$ and $i\in\llbracket 1,k\rrbracket$. We define a first empirical probability based on $n$ observations, following equation~\eqref{eq:probaempirique}, and a second empirical probability based on $m$ other independent observations,
$$\widehat q_{m,i}=\frac{1}{m}\sum_{j=n+1}^{n+m}\indic_{X_j\in\Omega_i}.$$
In what follows, we study both $D_{\text{KL}}\left(\widehat{\mathbf p}_n\Vert \mathbf p\right)$ and $D_{\text{KL}}\left(\widehat{\mathbf p}_n\Vert \widehat{\mathbf q}_m\right)$. It is worth mentioning a specific challenge in the two-sample relative entropy: even though $p_i>0$ whatever $i$, one cannot guarantee that $\widehat q_{m,i}\neq 0$. Beyond this trivial situation which leads to an infinite estimate, $\widehat q_{m,i}$ can be lower than the true value $p_i$ and amplify much the empirical relative entropy.

\section{Asymptotic and pre-asymptotic distributions of empirical relative entropy}\label{sec:CLT}

We first focus on the asymptotic distribution of empirical relative entropy, with one or two samples, thanks to the central limit theorem. Then, an extension of Berry-Esseen bounds to a nonlinear function of a sum statistic provides us with a non-asymptotic expression converging in distribution toward the limit of the central limit theorem. We are assuming that the data are generated according to the probability $\mathbf p$, so that the theoretical relative entropy should be equal to zero. Because of a well-known bias, the empirical relative entropy is positive.

\subsection{One-sample case}

The main challenge, when studying the statistical properties of the empirical relative entropy, is that we have a nonlinear function of the observations. A Taylor expansion can however make the problem feasible. Unlike what is done in the classical delta method, the first-order term of the expansion is equal to zero, so we need a second-order expansion and thus a convergence toward a chi-squared distribution~\cite{Bavaud,MJT}. Another possibility consists in using Wilks' theorem~\cite{Wilks}.

Regarding the speed of convergence toward the chi-squared distribution, we would like to use a Berry-Esseen approach. However, the literature dedicated to Berry-Esseen pre-asymptotic bounds in the case of a nonlinear statistic is very recent and still narrow. The purpose of Berry-Esseen inequality is to provide an upper bound to the Kolmogorov-Smirnov statistic between the distribution of a finite-sample statistic and its limit according to the central limit theorem. When the statistic is a nonlinear function of the observations, it may be possible to linearise it and thus to express the divergence with respect to a Gaussian distribution~\cite{PM,SZ}. This solution is not relevant in our case because relative entropy requires at least a quadratic approximation and has a non-Gaussian limit. Divergence with respect to non-Gaussian distributions have been scarcely explored in the literature. One can cite a first attempt with a statistic equal to the square of the sum of the observations, the limit being $\chi^2_1$~\cite{GT}. It is a first step but it is not enough in our case for which the limit is $\chi^2_{k-1}$. Promising results have been obtained in multidimensional extensions, with a chi-square limit, but with a number of degrees of freedom higher than 9~\cite{BG} or 5~\cite{GZ2011,GZ2014}, and a constant in the bound not explicitly specified or obtained by an indirect numerical procedure, requiring for example the number of integer vectors in a given ellipsoid~\cite[Theorem 1]{GU}. A very recent article also puts forward a solution which is valid whatever the number of degrees of freedom of the chi-squared distribution, but with unspecified constants and a limited domain of validity that excludes the right tail of the distribution~\cite{FLS}. Unfortunately, the right tail is quite important when the motivation is to build a statistical test of divergence.

Theorem~\ref{thm:TCL} gives the central limit of the one-sample relative entropy along with pre-asymptotic bounds in a Berry-Esseen approach. These bounds constitute a new result. We think it is one of the very rare attempts to obtain Berry-Esseen bounds for a statistic defined by a non-trivial nonlinear function of observations which does not converge to a Gaussian. It takes into account both the error of the quadratic approximation, known as the relative Pearson divergence, and the speed of convergence of this approximation. It exploits Rai\v{c}'s theorem, which is a recent multivariate extension of Berry-Esseen inequality, with well-specified constants~\cite{Raic}. 

\begin{thm}\label{thm:TCL}
Let $X_1,...,X_n$ be iid variables such that $\proba(X_j\in\Omega_i)=p_i$ with $\mathbf p=(p_1,...,p_k)^t\in(0,1)^k$. Then, when $n\rightarrow\infty$, we have
\begin{equation}\label{eq:asympt_KL}
\boxed{2nD_{\text{KL}}\left(\widehat{\mathbf p}_n\Vert \mathbf p\right) \overset{\text{d}}{\longrightarrow} \chi^2_{k-1}},
\end{equation}
where $\overset{\text{d}}{\longrightarrow}$ stands for the convergence in distribution. Let $x> 0$. We have
\begin{equation}\label{eq:BerryEsseen_KL}
\boxed{F_{\chi^2_{k-1}}\left(\kappa_{n,k}^{\text{down}}(x)\right) - \mathcal E_{n,k} \leq \proba\left(2nD_{\text{KL}}\left(\widehat{\mathbf p}_n\Vert \mathbf p\right)\leq x\right) \leq F_{\chi^2_{k-1}}\left(\kappa_{n,k}^{\text{up}}(x)\right) + \mathcal E_{n,k}},
\end{equation}
where $F_{\chi^2_{k-1}}$ is the cdf of the $\chi^2_{k-1}$ distribution,
\begin{equation}\label{eq:mathcalE}
\mathcal E_{n,k}=\left(42(k-1)^{1/4}+16\right)\sum_{i=1}^k \frac{(1-p_i)^{3/2}}{(np_i)^{1/2}},
\end{equation}
and where, for $\eta\in\{\text{up},\text{down}\}$, we have
$$\sqrt{\kappa_{n,k}^{\eta}(x)}=\left\{\begin{array}{ll}
\min\{(-1)^{\indic_{\eta=\text{up}}}\kappa_{n,k,r}(x)|r\in\{0,1,2\},(-1)^{\indic_{\eta=\text{up}}}\kappa_{n,k,r}(x)>0\} & \text{if } 27x\leq 4\mu n \\
\min\{(-1)^{\indic_{\eta=\text{up}}}\kappa_{n,k,>}(x)|(-1)^{\indic_{\eta=\text{up}}}\kappa_{n,k,>}(x)>0\} & \text{else,}
\end{array}\right.$$
with the convention $\min(\emptyset)=+\infty$, the notation $\mu=\min_{i\in\llbracket 1,k\rrbracket}p_i$, as well as
$$\kappa_{n,k,r}(x)=\frac{\sqrt{\mu n}}{3}\left[2\cos\left(\frac{1}{3}\arccos\left(\frac{27x}{2\mu n}-1\right)-\frac{2r\pi}{3}\right)-1\right]$$
and 
$$\kappa_{n,k,>}(x)=\sqrt[3]{\frac{\sqrt{\mu n}}{3}}\left[\sqrt[3]{ -\frac{\mu n}{9} + \frac{3x}{2} + \sqrt{-\frac{\mu n x}{3} +\frac{9}{4}x^2}}
+ \sqrt[3]{ -\frac{\mu n}{9} + \frac{3x}{2} - \sqrt{-\frac{\mu n x}{3} +\frac{9}{4}x^2}}\right]
- \frac{\sqrt{\mu n}}{3}.$$
Moreover, if $n\rightarrow\infty$, we have
\begin{equation}\label{eq:DL_kappa_up_down}
\left\{\begin{array}{ccl}
\kappa_{n,k}^{\text{up}}(x) & = & x+\frac{x^{3/2}}{\sqrt{\mu n}}+\mathcal O\left(\frac{1}{n}\right) \\
\kappa_{n,k}^{\text{down}}(x) & = & x-\frac{x^{3/2}}{\sqrt{\mu n}}+\mathcal O\left(\frac{1}{n}\right).
\end{array}\right.
\end{equation}
\end{thm}

The proof of Theorem~\ref{thm:TCL}, which is the most demanding of this work, follows in Section~\ref{sec:appendix_thm_TCL}.

Formula~\eqref{eq:BerryEsseen_KL} gives pre-asymptotic bounds for the cdf of the empirical relative entropy. The bounds deal with two approximations. First, $\mathcal E_{n,k}$ is the Berry-Esseen component, related to the speed of convergence toward the asymptotic distribution. Second, the chi-squared cdf are not simple functions of $x$ as it would be the case if relative entropy was a simple quadratic function. The variable $x$ is to be replaced by $\kappa_{n,k}^{\text{up}}(x)$ and $\kappa_{n,k}^{\text{down}}(x)$, which take into account the error of the quadratic approximation of relative entropy. These quantities are defined as the smallest positive solutions of a cubic equation. When $n$ increases, $\kappa_{n,k}^{\text{up}}(x)$ and $\kappa_{n,k}^{\text{down}}(x)$ tend toward $x$, as one can see in equation~\eqref{eq:DL_kappa_up_down}.

We think our approach can be reproduced for obtaining pre-asymptotic bounds for the cdf of some nonlinear function of sum statistics, under some convexity condition: considering a Taylor expansion of the nonlinear function, using Rai\v{c}'s theorem to get the speed of convergence, taking into account a bound of the residual, which in our case is more subtle than the maximum third derivative, which is infinite.

The Berry-Esseen part of the bounds is uniform in $x$. There is a recent effort in the literature to obtain non-uniform bounds in the linear framework~\cite{Pinelis}. Our problem could certainly benefit in the future from potential extensions of these non-uniform bounds to nonlinear functions of sum statistics.

In order to have an idea of the minimum $n$ making our Berry-Esseen bounds relevant, we consider the case of a uniform distribution for both $p$ and $q$. A computation of the quantities displayed in Theorem~\ref{thm:TCL} shows that $\mathcal E_{n,k}<1$ only when $n>1.37\times 10^5$ (respectively $n>3.36\times 10^3$) in the case $k=4$ (resp. $k=2$), underlining that the bounds are more pre-asymptotic than finite-sample. One may also wonder about the role of $\kappa_{n,k}^{\text{down}}(x)$ and $\kappa_{n,k}^{\text{up}}(x)$ in inequality~\ref{eq:BerryEsseen_KL}. Their importance is very limited because they converge rapidly toward $x$ as $n$ goes to infinity. Specifically, the differences $x- \kappa_{n,k}^{\text{down}}(x)$ and $\kappa_{n,k}^{\text{up}}(x)-x$ are increasing in $x$ and reach $0.32\%$ (respectively $0.45\%$, $0.64\%$, $0.91\%$) of $x$, for $x=10$, $n=2,000,000$, and $k=2$ (resp. $n=2,000,000$ and $k=4$, $n=500,000$ and $k=2$, $n=500,000$ and $k=4$).

\subsection{Proof of Theorem~\ref{thm:TCL}}\label{sec:appendix_thm_TCL}

The proof of Theorem~\ref{thm:TCL} uses Lemma~\ref{lem:cubicEquation}, which is provided in Appendix~\ref{sec:lem_cubicEquation}.

\begin{proof}
The observations $X_j$ being independent of each other, the random vector $n\widehat{\mathbf p}_n$ is a multinomial variable of parameter $\mathbf p$. The multivariate central limit theorem thus gives
$$\sqrt{n}\left(\widehat{\mathbf p}_n-\mathbf p\right)\overset{\text{d}}{\longrightarrow} \mathcal N\left(0,\diag(\mathbf p)-\mathbf p\mathbf p^t\right),$$
where $\mathcal N(\bm{\mu},\bm{\sigma}^2)$ is the Gaussian distribution of mean $\bm{\mu}$ and variance $\bm{\sigma}^2$. As a consequence, 
\begin{equation}\label{eq:TCL_vectorDiffProba}
\sqrt{n}\left(\begin{array}{c}
\frac{\widehat p_{n,1}-p_1}{\sqrt{p_1}} \\
\vdots \\
\frac{\widehat p_{n,k}-p_k}{\sqrt{p_k}}\end{array}\right) \overset{\text{d}}{\longrightarrow} \mathcal N\left(0,\bm{\Gamma}_k\right),
\end{equation}
where $\bm{\Gamma}_k=\mathbf I_k-\mathbf u\mathbf u^t$, with $\mathbf I_k$ the identity matrix and $\mathbf u=(\sqrt{p_1},...,\sqrt{p_k})^t$. We now need the eigenspaces of $\bm{\Gamma}_k$. Since $\bm{\Gamma}_k$ is real and symmetric, its eigenvectors form an orthogonal basis. First, $\mathbf u$ is an eigenvector associated with the eigenvalue 0:
$$\bm{\Gamma}_k \mathbf u=\mathbf u-\mathbf u\mathbf u^t\mathbf u=(1-\mathbf u^t\mathbf u)\mathbf u=\left(1-\sum_{i=1}^kp_i\right)\mathbf u=0.$$
Let's consider any vector $\mathbf v$ orthogonal to $\mathbf u$, that is $\mathbf u^t\mathbf v=0$. The space of all possible vectors $\mathbf v$ is thus of dimension $k-1$. Then, from the orthogonality condition, we easily get $\bm{\Gamma}_k \mathbf v=\mathbf v$. Therefore, the eigenspace of $\bm{\Gamma}_k$ associated with the eigenvalue 1 is of dimension $k-1$, which is also the rank of the matrix $\bm{\Gamma}_k$. The nonzero eigenvalues of $\bm{\Gamma}_k$ being all equal to 1, we have the reduced spectral decomposition $\bm{\Gamma}_k=\mathbf V\mathbf I_{k-1}\mathbf V^t=\mathbf V\mathbf V^t$, where $\mathbf V\in\mathbb R^{k\times (k-1)}$ is a matrix whose columns are orthonormal unit-eigenvalue eigenvectors, so that $\mathbf V^t\mathbf V=\mathbf I_{k-1}$.

Let us now focus on the empirical relative entropy, $f(\widehat{\mathbf p}_n)$, written as a function of $\widehat{\mathbf p}_n$, where $f(\mathbf q)=D_{\text{KL}}\left(\mathbf q\Vert \mathbf p\right)$. The function $f$ can be decomposed in a sum of $k$ univariate functions: $f(\mathbf q)=\sum_{i=1}^kf_i(q_i)$, where $f_i(q_i)=q_i\log(q_i/p_i)$, $f_i'(q_i)=1+\log(q_i/p_i)$, and $f_i^{(j)}(q_i)=(-1)^j(j-2)!q_i^{1-j}$ for $j\geq 2$. This decomposition eases the second-order Taylor expansion of $f$, which can be seen as the sum of $k$ distinct expansions, and where we also note that $f_i(p_i)=0$ and $f_i'(p_i)=1$:
\begin{equation}\label{eq:Taylor_KL}
f(\widehat{\mathbf p}_n)=\frac{1}{2}\sum_{i=1}^k\frac{(\widehat p_{n,i}-p_i)^2}{p_i}+\sum_{i=1}^k R_i(\widehat p_{n,i}),
\end{equation}
with $R_i(\widehat p_{n,i})=f_i^{(3)}(\xi_{n,i})(\widehat p_{n,i}-p_i)^3/6=(p_i-\widehat p_{n,i})^3/6\xi_{n,i}^2$, where $\xi_{n,i}$ is in the interval delimited by $\widehat p_{n,i}$ and $p_i$. 

From the decomposition $\bm{\Gamma}_k=\mathbf V\mathbf V^t$, we can write the limit appearing in formula~\eqref{eq:TCL_vectorDiffProba} as $\mathbf V\mathbf G$, where $\mathbf G\in\mathbb R^{k-1}$ is a standard Gaussian vector. Then,
$$\sum_{i=1}^k\frac{n(\widehat p_{n,i}-p_i)^2}{p_i} \overset{\text{d}}{\longrightarrow} \mathbf G^t\mathbf V^t\mathbf V\mathbf G=\mathbf G^t\mathbf G,$$
using the orthonormal property of $\mathbf V$. This limit follows a chi-square distribution $\chi^2_{k-1}$. Moreover, since $\widehat p_{n,i}\overset{\proba}{\longrightarrow}p_i$, we have $\xi_{n,i}\overset{\proba}{\longrightarrow}p_i$ and, by the continuous mapping theorem, $(p_i-\widehat p_{n,i})/6\xi_{n,i}^2\overset{\proba}{\longrightarrow}0$, because $p_i\neq 0$. Therefore, $nR_i(\widehat p_{n,i})=n(p_i-\widehat p_{n,i})^2(p_i-\widehat p_{n,i})/6\xi_{n,i}^2\overset{\proba}{\longrightarrow} 0$, since it is a product of a sequence converging in distribution with a sequence converging in probability to zero. Finally, starting from equation~\eqref{eq:Taylor_KL}, we see that $2nf(\widehat{\mathbf p}_n)$ is a sum of a sequence converging in distribution to $\chi^2_{k-1}$ with $k$ sequences converging in probability to zero. Slutsky's theorem thus leads to equation~\eqref{eq:asympt_KL}.

We now have to prove the second part of the theorem. We're going to use a multivariate extension of Berry-Esseen theorem~\cite{Raic} and apply it first to the quadratic part of $f$: the rest, namely the $R_i$ functions, will be considered in a second stage. We define $\mathbf Y_1$, ..., $\mathbf Y_n$, iid vectors of $\mathbb R^k$, by
$$\mathbf Y_j=\left(\frac{\indic_{X_j\in\Omega_1}-p_1}{\sqrt{np_1}},...,\frac{\indic_{X_j\in\Omega_k}-p_k}{\sqrt{np_k}}\right)^t.$$
It is easy to see that $\E(\mathbf Y_j)=0$. We will need the expression of the third moment of the Euclidean norm of $\mathbf Y_j$:
\begin{equation}\label{eq:moment3_normeEucl}
\begin{array}{ccl}
\E\left[\left(\mathbf Y_j^t\mathbf Y_j\right)^{3/2}\right] & = & \E\left[\left(\sum_{i=1}^k\frac{(\indic_{X_j\in\Omega_i}-p_i)^2}{np_i}\right)^{3/2}\right] \\
 & = & \sum_{u=1}^k\proba(X_j\in\Omega_u)\left(\frac{(1-p_u)^2}{np_u}+\sum_{i\neq u}\frac{p_i^2}{np_i}\right)^{3/2} \\
 & = & \frac{1}{n^{3/2}}\sum_{u=1}^k p_u\left(\frac{(1-p_u)^2}{p_u}+1-p_u\right)^{3/2} \\
 & = & \frac{1}{n^{3/2}}\sum_{u=1}^k \frac{(1-p_u)^{3/2}}{p_u^{1/2}}. 
\end{array}
\end{equation}
We also define $\mathbf W=\sum_{j=1}^n\mathbf Y_j$, whose $i$-th component, for $i\in\llbracket 1,k\rrbracket$, is $(\widehat p_{n,i}-p_i)(n/p_i)^{1/2}$, and 
$$\Theta=\mathbf W^t\mathbf W=\sum_{i=1}^k\frac{n(\widehat p_{n,i}-p_i)^2}{p_i}.$$
The properties of multinomial variables indicate that the covariance matrix of the vector $\mathbf W$ is $\bm{\Gamma}_k=\mathbf V\mathbf V^t$, whose rank is $k-1$ as explained above. Therefore, we can decompose $\mathbf W$ in an orthonormal basis of dimension $k-1$: we define a vector $\mathbf U\in\mathbb R^{k-1}$ as $\mathbf V^t\mathbf W$, so that $\mathbf W=\mathbf V\mathbf U$ and $\Theta=\mathbf U^t\mathbf U$. Then, it appears that $\mathbf U=\sum_{j=1}^n \mathbf T_j$, where $\mathbf T_j=\mathbf V^t\mathbf Y_j$. By linearity, $\E(\mathbf T_j)=0$ and, using the independence of $\mathbf Y_j$ and $\mathbf Y_{\ell}$ for $\ell\neq j$, we have as well
$$\begin{array}{ccl}
\sum_{j=1}^n\E(\mathbf T_j\mathbf T_j^t) & = & \sum_{j=1}^n\mathbf V^t\E(\mathbf Y_j\mathbf Y_j^t)\mathbf V \\
 & = & \mathbf V^t\E(\mathbf W\mathbf W^t)\mathbf V \\
 & = & \mathbf V^t\bm{\Gamma}_k \mathbf V \\
 & = & \mathbf V^t\mathbf V\mathbf V^t\mathbf V=\mathbf I_{k-1}.
\end{array}$$
These conditions on $\mathbf U$ and $\mathbf T_j$ are the ones that are required for Rai\v{c}'s multivariate Berry-Esseen theorem~\cite[Theorem 1.1]{Raic}. For a proper application of this theorem, two other assumptions are still to be verified. First, because $\mathbf Y_j=\mathbf V\mathbf T_j$, $\mathbf Y_j$ and $\mathbf T_j$ have the same Euclidean norm, $\mathbf Y_j^t\mathbf Y_j=\mathbf T_j^t\mathbf V^t\mathbf V\mathbf T_j=\mathbf T_j^t\mathbf T_j$, so that $\E\left[\left(\mathbf T_j^t\mathbf T_j\right)^{3/2}\right]=\sum_{u=1}^k (1-p_u)^{3/2}/n^{3/2}p_u^{1/2}$ after equation~\eqref{eq:moment3_normeEucl}. Second, the set $\mathcal A_{k-1,x}=\{\mathbf Z\in\mathbb R^{k-1}|\mathbf Z^t\mathbf Z\leq x\}$ is convex because it is the sublevel set of a convex function. We now have all the conditions for applying Rai\v{c}'s theorem~\cite[Theorem 1.1]{Raic}:
\begin{equation}\label{eq:Raic_sanserreur}
\begin{array}{ccl}
\left|\proba\left(\Theta\leq x\right)-F_{\chi^2_{k-1}}(x)\right| & = & \left|\proba\left(\mathbf U\in\mathcal A_{k-1,x}\right)-\proba(\mathbf G\in\mathcal A_{k-1,x})\right| \\
 & \leq & \left(42(k-1)^{1/4}+16\right)\sum_{j=1}^n\E\left[\left(\mathbf T_j^t\mathbf T_j\right)^{3/2}\right] \\
 & \leq & \left(42(k-1)^{1/4}+16\right)\sum_{u=1}^k \frac{(1-p_u)^{3/2}}{(np_u)^{1/2}},
 \end{array}
\end{equation}
where $\mathbf G\in\mathbb R^{k-1}$ is a standard Gaussian vector. 

We now adapt the above Berry-Esseen inequality to include the residual $\mathcal R=2n\sum_{i=1}^k R_i(\widehat p_{n,i})$ of $2nf(\widehat{\mathbf p}_{n})$, given by the Taylor expansion displayed in equation~\ref{eq:Taylor_KL}. If $\widehat p_{n,i}\geq p_i$, then $\xi_{n,i}\geq p_i$ and we simply have $|R_i(\widehat p_{n,i})|\leq|p_i-\widehat p_{n,i}|^3/6p_i^2$. If $\widehat p_{n,i}< p_i$, we cannot properly bound the residual of the second-order Taylor expansion with the same method and we need a more precise analysis. The non-truncated Taylor series provides, for $x\in[0,p]$,
$$R_i(x)=(x-p_i)^3\sum_{j=3}^\infty\frac{(-1)^j(x-p_i)^{j-3}}{j(j-1)p_i^{j-1}}.$$
Noting $h_i(x)=R_i(x)/(x-p_i)^3$, we have
$$h'_i(x)=\sum_{j=3}^\infty\frac{(p_i-x)^{j-2}(j-3)}{j(j-1)p_i^{j-1}},$$
which is non-negative whatever $x\leq p_i$. Since we also have $h_i(p_i)=-1/6p_i^2$ and $h_i(0)=-1/2p_i^2$, we finally get $\max_{x\in[0,p_i]}|h_i(x)|=1/2 p_i^2$. Combining this result with our analysis for $x\geq p_i$, we find that $1/2p_i^2$ is an upper bound for $|h(x)|$ whatever $x\in[0,1]$. Then, noting that $L^q$ norms are non-increasing functions of $q$, and writing $\mu=\min_{i\in\llbracket 1,k\rrbracket}p_i$, we get
$$\left|\mathcal R\right| \leq 2n\sum_{i=1}^k \frac{\left|p_i-\widehat p_{n,i}\right|^3}{2p_i^2} \leq \frac{n}{\sqrt{\mu}}\left(\sum_{i=1}^k \left|\frac{p_i-\widehat p_{n,i}}{\sqrt{p_i}}\right|^2\right)^{3/2} = \frac{\Theta^{3/2}}{\sqrt{\mu n}}.$$
Since $2nf(\widehat{\mathbf p}_{n})=\Theta+\mathcal R$ and $(\Theta\leq x-\rho) \Rightarrow (\Theta+\mathcal R\leq x) \Rightarrow (\Theta\leq x+\rho)$ for $\rho\geq |\mathcal R|$, we have
\begin{equation}\label{eq:K_down_up_first}
\proba\left(K_{n,k}^{\text{down}}(\Theta)\leq x\right) \leq \proba\left(2nf(\widehat{\mathbf p}_{n})\leq x\right) \leq \proba\left(K_{n,k}^{\text{up}}(\Theta)\leq x\right),
\end{equation}
where $K_{n,k}^{\text{up}}:z\geq 0\mapsto z-(\mu n)^{-1/2} z^{3/2}$ and $K_{n,k}^{\text{down}}:z\geq 0\mapsto z+(\mu n)^{-1/2} z^{3/2}$. Also, because of the convexity of $f$~\cite[Theorem 2.7.2]{CT}, we have the convexity of the set $\{\mathbf q|2nf(\mathbf q)\leq x\}$. So we can refine formula~\eqref{eq:K_down_up_first} by taking into account a constraint of convexity, that is the three sets in this formula have to be convex. Noting that $K_{n,k}^{\text{up}}(0)=K_{n,k}^{\text{down}}(0)=0< x$ and that $(\Theta=0)\Rightarrow(\forall i, \widehat p_{n,i}=p_i)\Rightarrow (2nf(\widehat{\mathbf p}_{n})=0< x)$, 0 belongs to the three intervals of admissible values of $\Theta$ and the convexity constraint modifies inequalities~\eqref{eq:K_down_up_first} in
$$\proba\left(\Theta\in[0,\kappa_{n,k}^{\text{down}}(x)]\right) \leq \proba\left(2nf(\widehat{\mathbf p}_{n})\leq x\right) \leq \proba\left(\Theta\in[0,\kappa_{n,k}^{\text{up}}(x)]\right),$$
where $\kappa_{n,k}^{\text{up}}(x)$ (respectively $\kappa_{n,k}^{\text{down}}(x)$) is the smallest positive root of $z\mapsto K_{n,k}^{\text{up}}(z)-x$ (resp. $z\mapsto K_{n,k}^{\text{down}}(z)-x$) if it exists, otherwise the probabilities are trivially equal to 1. An explicit expression for $\kappa_{n,k}^{\text{up}}(x)$ and $\kappa_{n,k}^{\text{down}}(x)$ is obtained as the solution of a cubic equation, solved with Cardano's formula. Indeed, using Lemma~\ref{lem:cubicEquation} applied to $\sqrt{z}$, with $d=x$ and $a=-(\mu n)^{-1/2}$ for $\kappa_{n,k}^{\text{up}}(x)$ or $a=(\mu n)^{-1/2}$ for $\kappa_{n,k}^{\text{down}}(x)$, we directly obtain the expression displayed in Theorem~\ref{thm:TCL}. We also have $(\mu n)^{-1/2}\rightarrow 0$ and Lemma~\ref{lem:cubicEquation} provides us with the asymptotic behaviour of $\kappa_{n,k}^{\text{up}}(x)$, namely 
$$\sqrt{\kappa_{n,k}^{\text{up}}(x)}=\sqrt{x}+\frac{x}{2\sqrt{\mu n}}+\mathcal O\left(\frac{1}{n}\right),$$
which gives
$$\kappa_{n,k}^{\text{up}}(x)=x+\frac{x^{3/2}}{\sqrt{\mu n}}+\mathcal O\left(\frac{1}{n}\right).$$
A similar idea makes it possible to lead to the result displayed in Theorem~\ref{thm:TCL} for $\kappa_{n,k}^{\text{down}}(x)$.

Finally, we can use Rai\v{c}'s theorem again, using directly formula~\eqref{eq:Raic_sanserreur} with a new value for $x$:
$$\begin{array}{ccl}
\proba\left(2nf(\widehat{\mathbf p}_{n})\leq x\right) & \leq & \proba\left(\Theta\leq\kappa_{n,k}^{\text{up}}(x)\right) \\
 & \leq & F_{\chi^2_{k-1}}(\kappa_{n,k}^{\text{up}}(x)) + \left|\proba\left(\Theta\leq \kappa_{n,k}^{\text{up}}(x)\right)-F_{\chi^2_{k-1}}(\kappa_{n,k}^{\text{up}}(x))\right| \\
 & \leq & F_{\chi^2_{k-1}}(\kappa_{n,k}^{\text{up}}(x)) + \left(42(k-1)^{1/4}+16\right)\sum_{u=1}^k \frac{(1-p_u)^{3/2}}{(np_u)^{1/2}}.
\end{array}$$
We similarly get
$$\begin{array}{ccl}
\proba\left(2nf(\widehat{\mathbf p}_{n})\leq x\right) & \geq & \proba\left(\Theta\leq\kappa_{n,k}^{\text{down}}(x)\right) \\
 & \geq & F_{\chi^2_{k-1}}(\kappa_{n,k}^{\text{down}}(x)) - \left|\proba\left(\Theta\leq \kappa_{n,k}^{\text{down}}(x)\right)-F_{\chi^2_{k-1}}(\kappa_{n,k}^{\text{up}}(x))\right|
\end{array}$$
and we can now conclude with the statement of the theorem.
\end{proof}

\subsection{Two-sample case}

For the two-sample problem, Theorem~\ref{thm:TCL_two-sample} proposes an asymptotic distribution. We haven't found such a result in the literature but it is worth mentioning a close contribution with the asymptotic distribution of the two-sample Jeffreys divergence, which is a symmetric version of relative entropy~\cite{GLL}.

\begin{thm}\label{thm:TCL_two-sample}
Let $X_1,...,X_{n+m}$ be iid variables such that $\proba(X_j\in\Omega_i)=p_i$ with $\mathbf p=(p_1,...,p_k)^t\in(0,1)^k$. Then, when $n\rightarrow\infty$, $m\rightarrow\infty$, and $\frac{n}{n+m}\rightarrow\lambda\in(0,1)$, we have
$$2\frac{nm}{n+m}D_{KL}(\widehat{\mathbf p}_{n}\Vert\widehat{\mathbf q}_{m}) \overset{\text{d}}{\longrightarrow} \chi^2_{k-1}.$$
\end{thm}

The proof of Theorem~\ref{thm:TCL_two-sample} is postponed in Appendix~\ref{sec:appendix_thm_TCL_two-sample}.

When the two samples have the same size, that is $n=m$, $D_{KL}(\widehat{\mathbf p}_{n}\Vert \widehat{\mathbf q}_{m})$ is asymptotically distributed like $\chi^2_{k-1}/n$. It is to be compared to the more concentrated asymptotic distribution of $D_{KL}(\widehat{\mathbf p}_{n}\Vert \mathbf p)$, which is $\chi^2_{k-1}/2n$.

In the two-sample case, we do not propose Berry-Esseen bounds. Indeed, in the one-sample case, we were able to find an upper bound of the rest expressed as a simple function of the two probabilities. But in the two-sample case the rest of the quadratic approximation of relative entropy depends on the divergence between each empirical probability and the true probability, that is $\widehat p_{n,i}-p_i$ and $\widehat q_{m,i}-p_i$, and not on the difference between the two empirical probabilities, $\widehat p_{n,i}-\widehat q_{m,i}$. We can however propose a Berry-Esseen bound for the quadratic approximation instead of the relative entropy itself, as shown in Proposition~\ref{pro:BerryEsseen_two-sample}. In Section~\ref{sec:concentration}, we will present finite-sample results in the two-sample case, directly applied to relative entropy.

\begin{pro}\label{pro:BerryEsseen_two-sample}
With the assumptions of Theorem~\ref{thm:TCL_two-sample} and $x>0$, we have
$$\left|\proba\left(\frac{nm}{n+m}\sum_{i=1}^k \frac{\left(\widehat p_i-\widehat q_i\right)^2}{p_i}\leq x\right)-F_{\chi^2_{k-1}}(x)\right| \leq \frac{n^2+m^2}{(nm)^{1/2}(n+m)} \mathcal E_{n+m,k},$$
with $\mathcal E_{.,.}$ defined in equation~\eqref{eq:mathcalE}.
\end{pro}

The proof of Proposition~\ref{pro:BerryEsseen_two-sample} is postponed in Appendix~\ref{sec:appendix_thm_TCL_two-sample}.

When $n=m$, the bound in Proposition~\ref{pro:BerryEsseen_two-sample} is $\mathcal E_{2n,k}$, that is $\mathcal E_{n,k}/2^{1/2}$.

\section{Concentration inequalities for empirical relative entropy}\label{sec:concentration}

Beside the Berry-Esseen bounds, one can obtain finite-sample bounds of the distribution of the empirical relative entropy by the mean of concentration inequalities. They in general offer simpler expressions than Berry-Esseen bounds, without referring to the limit distribution. Instead, they exploit various methods such as the method of types for the famous Sanov inequality, and a recursion technique or the moment-generating function in the two promising alternatives we present below.

In what follows, we expose three existing inequalities, with two among the most recent and promising ones, along with a small refinement for the last one. We propose as well new concentration inequalities in the two-sample case. We will finish by a short simulation study.

\subsection{One-sample case}

Sanov inequality is the most well-known concentration inequality for relative entropy. It writes
\begin{equation}\label{eq:Sanov}
\proba\left(D_{\text{KL}}\left(\widehat{\mathbf p}_n\Vert \mathbf p\right)\geq x\right) \leq \frac{(n+k-1)!}{n!(k-1)!}e^{-n x} \leq (n+1)^ke^{-n x},
\end{equation}
some works focusing on the first inequality~\cite{Csiszar,MJT}, while others prefer the second one~\cite[Theorem 11.2.1]{CT}, which is a simpler bound, in particular when $k$ is large.

Based on a recursion technique, Mardia's bounds improve Sanov inequality, in particular when one increases $k$. There are several Mardia's bounds, each of which apply to a specific range of values for $k$. Among them, we focus on the one that demonstrated superior performance for the values of $k$ considered in our tests:
\begin{equation}\label{eq:Mardia}
\proba\left(D_{\text{KL}}\left(\widehat{\mathbf p}_n\Vert \mathbf p\right)\geq x\right) \leq \frac{6e^2}{\pi^{3/2}}\left(\frac{ne^3}{2\pi k}\right)^{k/2} e^{-nx}.
\end{equation}
It holds when $3\leq k \leq 2+\sqrt{ne^3/2\pi}$~\cite{MJT}.

Agrawal proposed another concentration inequality exploiting the moment-generating function of the empirical relative entropy~\cite{Agrawal}. We expose it in the following proposition, along with a slightly improved version.

\begin{pro}\label{pro:Agrawal}
If $x>(k-1)/n$, then
$$\proba(D_{\text{KL}}\left(\widehat{\mathbf p}_n\Vert \mathbf p\right)\geq x) \leq \mathcal M^{1}_{k,n}(x) \leq \mathcal M^{2}_{k,n}(x) \leq \mathcal M^{3}_{k,n}(x),$$
with
$$\left\{\begin{array}{ccl}
\mathcal M^{1}_{k,n}(x) & = & \inf_{t\in[0,n)}e^{-tx}\left(\sum_{j=0}^n \frac{n!}{n^{2j}(n-j)!}t^j\right)^{k-1} \\
\mathcal M^{2}_{k,n}(x) & = & e^{-nx}\left(\sum_{j=0}^n \frac{n!e}{n^{j}(n-j)!}\left(1-\frac{k-1}{nx}\right)^j\right)^{k-1} \\
\mathcal M^{3}_{k,n}(x) & = & e^{-nx}\left(\frac{enx}{k-1}\right)^{k-1}.
\end{array}\right.$$
\end{pro}

The proof of Proposition~\ref{pro:Agrawal} is postponed in Appendix~\ref{sec:appendix_pro_Agrawal}.

The third bound in Proposition~\ref{pro:Agrawal} is the one put forward by Agrawal. As we will see in Section~\ref{sec:simulevalbounds}, it is both simple and very performing, compared to the ones of equations~\eqref{eq:Sanov} and~\eqref{eq:Mardia}, at least when $k$ is small. The second bound, $\mathcal M^{2}_{k,n}(x)$, slightly improves $\mathcal M^{3}_{k,n}(x)$, but it is more appropriate when $n$ is small because it requires the calculation of a sum of $n$ terms. The algorithmic complexity for obtaining the first bound, $\mathcal M^{1}_{k,n}(x)$, is even worse. Indeed, in addition to the sum of $n$ terms, it requires a numerical optimization.

We also remark, beyond the traditional $e^{-n x}$ of the concentration inequalities, that $x$ appears in other parts of the expression of the bounds of Proposition~\ref{pro:Agrawal}, whereas neither equation~\eqref{eq:Sanov} nor equation~\eqref{eq:Mardia} exhibits this. The consequence is that Agrawal's bounds are closer to the true probability when $x$ is small, compared to other methods. When one looks for a quantile at a given probability, $k$ small or $n$ not too small lead to a small quantile and thus Agrawal's formula provides us with a tighter upper bound of the quantile, compared to the alternatives of equations~\eqref{eq:Sanov} and~\eqref{eq:Mardia}. This will be confirmed in the simulations presented in Section~\ref{sec:simulevalbounds}.

\subsection{Two-sample case}

For building a concentration inequality in the two-sample framework, we can split relative entropy in two parts so that we can directly sum the upper bounds of the one-sample case. But such a decomposition does not naturally arise because there is no triangle inequality for relative entropy. Pinsker's inequality shows however a correspondence between relative entropy and total variation, which could be used to obtain the desired decomposition. For discrete probabilities, it writes
\begin{equation}\label{eq:Pinsker}
\left(\sum_{i=1}^k |p_i-q_i|\right)^2 \leq 2D_{\text{KL}}\left(\mathbf p\Vert \mathbf q\right) \leq \left(\frac{1}{\min_{i\in\llbracket 1,k\rrbracket} q_i}-\min_{i\in\llbracket 1,k\rrbracket} \frac{p_i}{q_i}\right) \left(\sum_{i=1}^k |p_i-q_i|\right)^2,
\end{equation}
the left bound being the traditional Pinsker's inequality~\cite{Tsybakov,Canonne} and the right bound one of the reverse Pinsker's inequalities~\cite{SV2015}, among several other versions~\cite{BH,SV2016}. We note that $(1/\min_i q_i)-(\min_i p_i/q_i) \geq k-1$. We thus get, as an interesting side result, the following decomposition of relative entropy.

\begin{pro}\label{pro:inegKL}
Let $\mathbf p$, $\mathbf q$, and $\mathbf r$ be categorical distributions with $k$ categories. If $\min_{i\in\llbracket 1,k\rrbracket} q_i>0$, we have:
$$D_{\text{KL}}\left(\mathbf p\Vert \mathbf q\right) \leq \left(\frac{2}{\min_{i\in\llbracket 1,k\rrbracket} q_i}-2\min_{i\in\llbracket 1,k\rrbracket} \frac{p_i}{q_i}\right) \left(D_{\text{KL}}\left(\mathbf p\Vert \mathbf r\right)+D_{\text{KL}}\left(\mathbf q\Vert \mathbf r\right)\right).$$
\end{pro}

The proof of Proposition~\ref{pro:inegKL} is postponed in Appendix~\ref{sec:appendix_inegKL}.

In the particular case $\mathbf r=\mathbf p$, it simplifies to
$$D_{\text{KL}}\left(\mathbf p\Vert \mathbf q\right) \leq \left(\frac{2}{\min_{i\in\llbracket 1,k\rrbracket} q_i}-2\min_{i\in\llbracket 1,k\rrbracket} \frac{p_i}{q_i}\right) D_{\text{KL}}\left(\mathbf q\Vert \mathbf p\right).$$
The scalar in front of the relative entropy on the right-hand side of the above equation can be very large when one considers a probability $\mathbf q$ that is far from being uniform and a probability $\mathbf p$ that largely diverges from $\mathbf q$. We also note that, in Proposition~\ref{pro:inegKL}, the probabilities can be exchanged in each relative entropy of the right-hand side of the inequality. Therefore, $D_{\text{KL}}\left(\mathbf p\Vert \mathbf r\right)+D_{\text{KL}}\left(\mathbf q\Vert \mathbf r\right)$ can for example be replaced by $D_{\text{KL}}\left(\mathbf p\Vert \mathbf r\right)+D_{\text{KL}}\left(\mathbf r\Vert \mathbf q\right)$.

Using the one-sample Agrawal's concentration inequality introduced in Proposition~\ref{pro:Agrawal} along with the decomposition put forward in Proposition~\ref{pro:inegKL}, we get a concentration inequality for the two-sample relative entropy, when the two samples are assumed to follow the same theoretical distribution. It is the purpose of Theorem~\ref{thm:Agrawal_twosample}. In this finite-sample framework, the probability that $\min_{i\in\llbracket 1,k\rrbracket} \widehat q_{m,i}=0$ is significant and one must bypass this situation that leads to infinite relative entropy by considering, instead of the empirical probability $\widehat{\mathbf q}_m$, the regularized Laplace estimator $\overline{\mathbf q}_m$, defined by $\overline q_{m,i}=(1+m\widehat q_{m,i})/(m+k)$.

\begin{thm}\label{thm:Agrawal_twosample}
Let $m,n>0$ and $X_1,...,X_{n+m}$ be iid variables such that $\proba(X_j\in\Omega_i)=p_i$ with $\mathbf p=(p_1,...,p_k)^t\in(0,1)^k$. We note $\beta_{k,m}=2(m+k)$, $\zeta_{k,m}=-(k+m)^{-1}\sum_{i=1}^k\log(kp_i)$, and we assume that $x\geq\beta_{k,m}(k-1)(m+n)/mn$. Then,
$$\boxed{\proba\left(D_{\text{KL}}\left(\widehat{\mathbf p}_n\Vert \widehat{\mathbf q}_m\right)\geq x\right) \leq \widetilde{\mathcal M}^1_{k,n,m}(x) \leq \widetilde{\mathcal M}^2_{k,n,m}(x) \leq \widetilde{\mathcal M}^3_{k,n,m}(x)},$$
with
$$\left\{\begin{array}{ccl}
\widetilde{\mathcal M}^1_{k,n,m}(x) & = & \inf_{s\in[0,\min(m+k,n))}e^{s(\zeta_{k,m}-x/\beta_{k,m})}\left(\sum_{i=0}^{m+k} \frac{(m+k)!}{(m+k)^{2i}(m+k-i)!}s^i\sum_{j=0}^n \frac{n!}{n^{2j}(n-j)!}s^j\right)^{k-1} \\
\widetilde{\mathcal M}^2_{k,n,m}(x) & = & e^{\sigma_{m,n,x}(\zeta_{k,m}-x/\beta_{k,m})}\left(\sum_{i=0}^{m+k} \frac{(m+k)!}{(m+k)^{2i}(m+k-i)!}\sigma_{m,n,x}^i\sum_{j=0}^n \frac{n!}{n^{2j}(n-j)!}\sigma_{m,n,x}^j\right)^{k-1} \\
\widetilde{\mathcal M}^3_{k,n,m}(x) & = & e^{\sigma_{m,n,x}(\zeta_{k,m}-x/\beta_{k,m})}\left(\left(1-\frac{\sigma_{m,n,x}}{m+k}\right)\left(1-\frac{\sigma_{m,n,x}}{n}\right)\right)^{1-k}
\end{array}\right.$$
and $$\sigma_{m,n,x}=\frac{n+m+k}{2}-\frac{\beta_{k,m}(k-1)}{x-\beta_{k,m}\zeta_{k,m}}-\sqrt{\frac{\beta_{k,m}^2(k-1)^2}{(x-\beta_{k,m}\zeta_{k,m})^2}+\frac{(m+k-n)^2}{4}}.$$
\end{thm}

The proof of Theorem~\ref{thm:Agrawal_twosample} is postponed in Appendix~\ref{sec:appendix_Agrawal_twosample}.

When $m+k=n$ and $x>\beta_{k,m}\zeta_{k,m}$, then $\sigma_{m,n,x}$ becomes $n-2\beta_{k,m}(k-1)/(x-\beta_{k,m}\zeta_{k,m})$, the bounds simplify, and we get, for example,
$$\widetilde{\mathcal M}^3_{k,n,n-k}(x) = e^{-n(x-\beta_{k,m}\zeta_{k,m})/\beta_{n,n}}\left(\frac{en(x-\beta_{k,m}\zeta_{k,m})}{2\beta_{k,m}(k-1)}\right)^{2(k-1)}$$
or 
$$\widetilde{\mathcal M}^2_{k,n,n-k}(x) = e^{-n(x-\beta_{k,m}\zeta_{k,m})/\beta_{k,m}}\left(e\sum_{i=0}^{n}\left(1-\frac{2\beta_{k,m}(k-1)}{nx}\right)^i\left[\prod_{j=0}^{i-1}\left(1-\frac{j}{n}\right)\right]\right)^{2(k-1)},$$
expression in which we have replaced the ratio of factorials by a product that is equal but easier to compute for large values of $n$.
We now show that the parameter $\beta_{k,m}$ in the above bounds is quite pessimistic. Fixing the value of $\beta_{k,m}$ to the constant value $\beta$, we note that, when $m\rightarrow\infty$, $\zeta_{k,m}$ goes to zero, $\sigma_{m,n,x}\rightarrow\sigma_{n,x}=n-\beta(k-1)/x$, and
$$\widetilde{\mathcal M}^3_{k,n,m}(x) \rightarrow e^{-nx/\beta}\left(\frac{enx}{\beta(k-1)}\right)^{k-1}.$$
We remark that this limit is different from the expression of $\mathcal M_{k,n}^3(x)$ provided for the one-sample case in Proposition~\ref{pro:Agrawal}. The reason is the presence of $\beta$, which should be 1 in order for the limit to match the one-sample case. It thus appears that the reverse Pinsker's inequality is quite pessimistic and that the $\beta_{k,m}$ of Theorem~\ref{thm:Agrawal_twosample} is too large. In Section~\ref{sec:simulevalbounds}, we show that replacing $\beta_{k,m}$ by 1 and $\zeta_{k,m}$ by 0 leads to upper bounds that are numerically satisfying. We note $\widetilde{\mathcal M}^{2,\star}_{k,n,m}(x)$ and $\widetilde{\mathcal M}^{3,\star}_{k,n,m}(x)$ these new quantities. Even though we cannot prove it, we conjecture that Theorem~\ref{thm:Agrawal_twosample} holds for a large range of probabilities with these modified parameters.

\subsection{A simulation-based evaluation of the concentration bounds}\label{sec:simulevalbounds}

In this paragraph, we consider a uniform distribution for both $p$ and $q$. Indeed, numerical experiments indicate that the distribution of empirical relative entropy obtained from simulations does not seem to be significantly affected by changes in the probabilities, provided that the number of categories remains unchanged and that we stay under the hypothesis $\mathbf p=\mathbf q$.

We focus on finite-sample distributions. We will let $k$ and $n$ vary around the baseline case $n=100$, $k=4$. We compare quantiles of the empirical relative entropy at $75\%$ and $95\%$, in the one- and two-sample cases, following various methods: the empirical quantiles obtained from 10,000 simulations, the quantile in the asymptotic distribution of Theorems~\ref{thm:TCL} and~\ref{thm:TCL_two-sample}, and the quantiles obtained from concentration inequalities. Specifically, in the one-sample case, the considered bounds are the two Sanov's bounds of equation~\eqref{eq:Sanov}, Mardia's bound provided in equation~\eqref{eq:Mardia}, and Agrawal's second and third bounds, $\mathcal M^2_{k,n}$ and $\mathcal M^3_{k,n}$, defined in Proposition~\ref{pro:Agrawal}. With two samples, the only bounds considered are the extensions of the above Agrawal's bounds, $\widetilde{\mathcal M}^{2,\star}_{k,n,n}$ and $\widetilde{\mathcal M}^{3,\star}_{k,n,n}$, defined after Theorem~\ref{thm:Agrawal_twosample}, with the same size for the two subsamples.

Some of the concentration inequalities are valid under some condition. Agrawal's condition $x>(k-1)/n$ is always verified in our results and we have to discard the case $k=2$ for Mardia's bound. In the two-sample case, we also constrain $k$ to be less than or equal to 8; otherwise, simulations sometimes lead to empty categories, and consequently to zero estimated probabilities $\widehat{\mathbf q}_n$, when $n=100$. This would cause the relative entropy to diverge to infinity.

\begin{figure}[htb]
	\centering
		\includegraphics[width=0.45\textwidth]{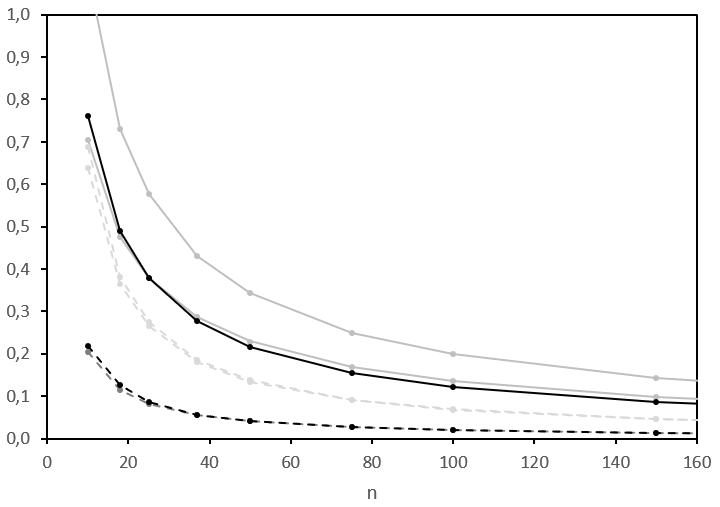} 
		\includegraphics[width=0.45\textwidth]{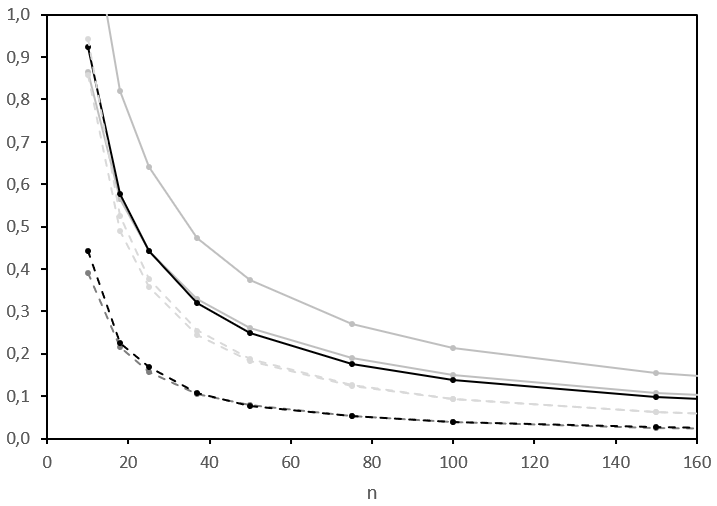}  \\
		\includegraphics[width=0.45\textwidth]{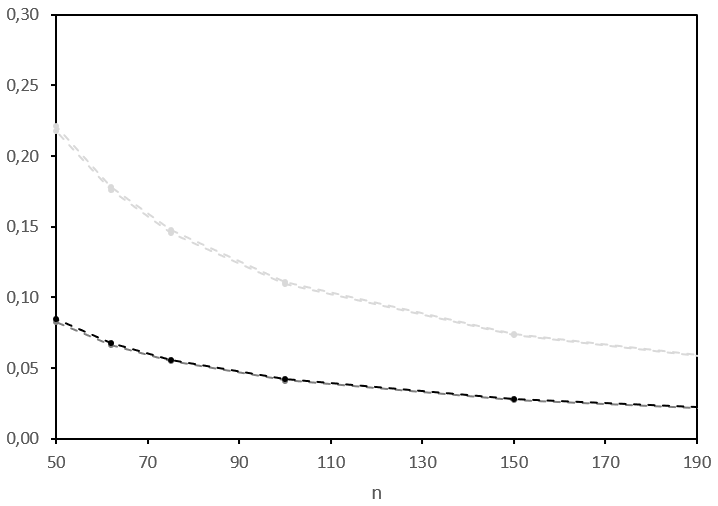} 
		\includegraphics[width=0.45\textwidth]{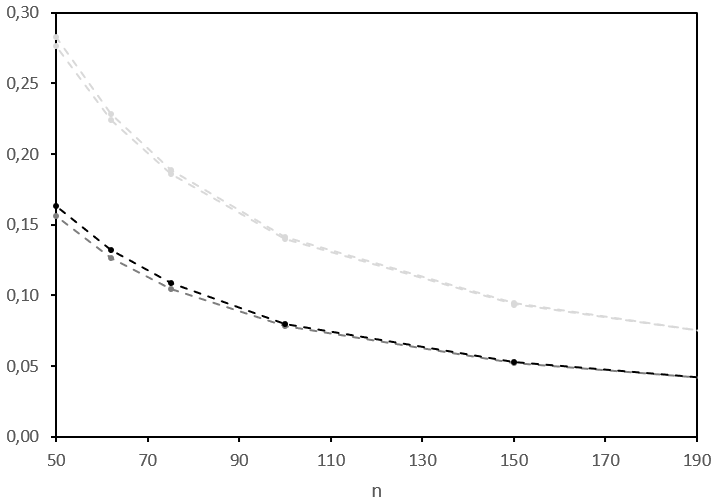} 
\begin{minipage}{0.7\textwidth}\caption{Quantile of the relative entropy as a function of $n$, at $75\%$ (left graphs) and $95\%$ (right graphs), for one (top graphs) or two samples (bottom graphs), according to the 10,000 simulations (black dashed line), to the asymptotic distribution (dark grey dashed line), to the second and third Agrawal's bounds (light grey dashed lines), to the two Sanov's bounds (grey solid lines), and to Mardia's bound (black solid line). Other parameters are $k=4$ and $p_i$ constant in $i$.}
	\label{fig:BornesSimulN}
\end{minipage}
\end{figure}

The obtained quantiles, as functions of $n$, are displayed in Figure~\ref{fig:BornesSimulN}. We observe that the quantile of the asymptotic distribution is very accurate: it only slightly underestimates the true quantile when $n$ is small (less than 50) and this underestimation becomes more pronounced the further we move into the tail of the distribution. Regarding concentration inequalities, Sanov's bounds largely overestimate the quantile. Mardia's bound is only slightly better than the best Sanov's bound, whereas the two Agrawal's bounds, which are very close to each other both in the one- and the two-sample cases, are significantly better than the other bounds, though not as accurate as the quantile of the asymptotic distribution.

\begin{figure}[htb]
	\centering
		\includegraphics[width=0.45\textwidth]{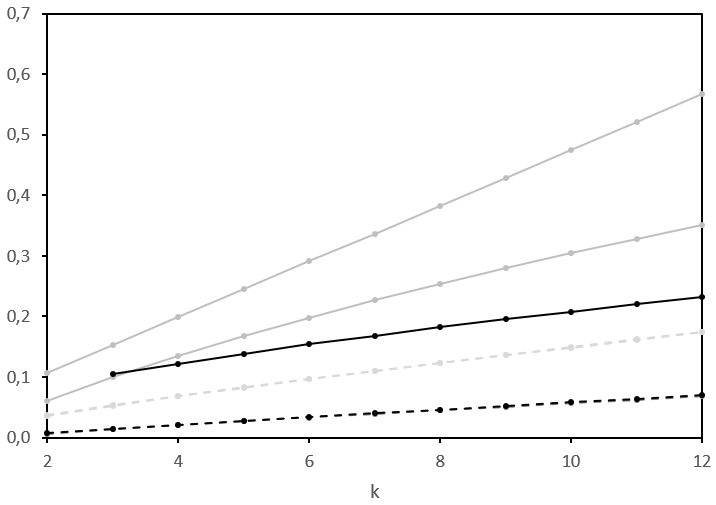} 
		\includegraphics[width=0.45\textwidth]{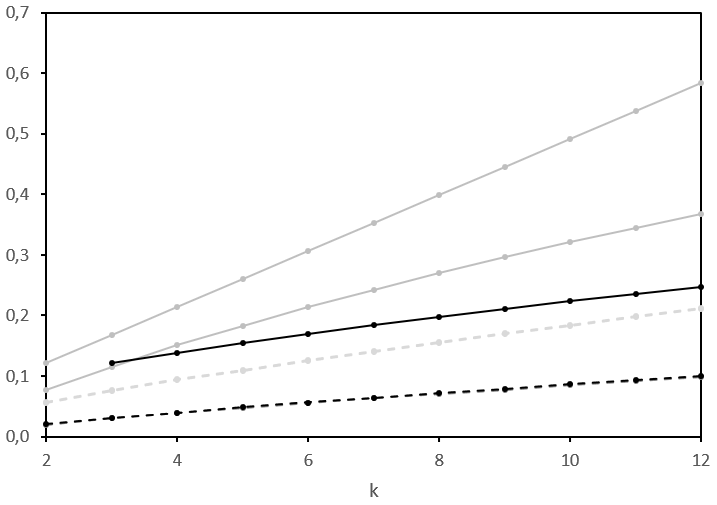}  \\
		\includegraphics[width=0.45\textwidth]{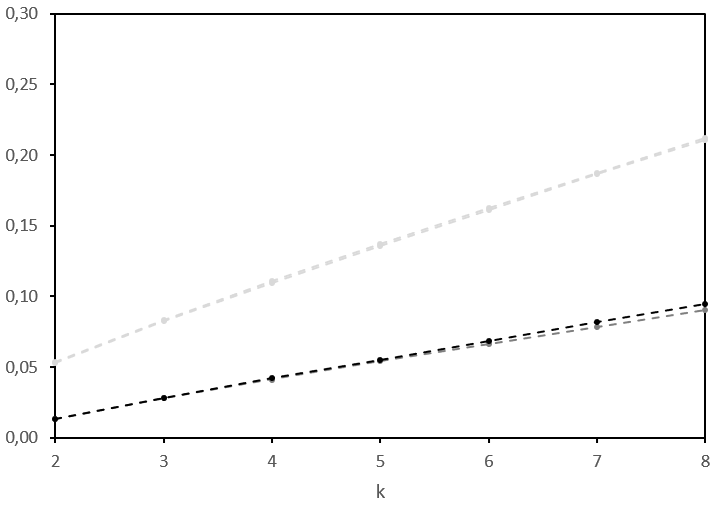} 
		\includegraphics[width=0.45\textwidth]{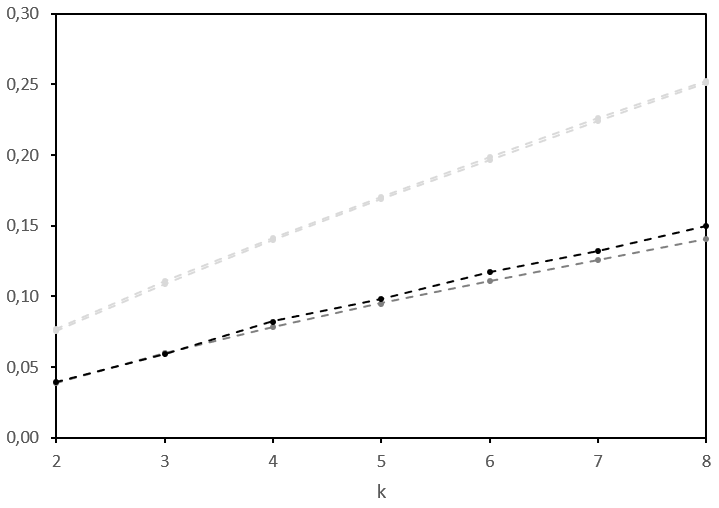} 
\begin{minipage}{0.7\textwidth}\caption{Quantile of the relative entropy as a function of $k$, at $75\%$ (left graphs) and $95\%$ (right graphs), for one (top graphs) or two samples (bottom graphs), according to the 10,000 simulations (black dashed line), to the asymptotic distribution (dark grey dashed line), to the second and third Agrawal's bounds (light grey dashed lines), to the two Sanov's bounds (grey solid lines), and to Mardia's bound (black solid line). Other parameters are $n=100$ and $p_i$ constant in $i$.}
	\label{fig:BornesSimulK}
\end{minipage}
\end{figure}

Figure~\ref{fig:BornesSimulK} shows the quantiles as functions of $k$. The quantile at $95\%$ of the asymptotic distribution more clearly underestimates the true quantile when $k$ is larger. Once again, Sanov's bounds are the least accurate. Mardia's bound comes next but it increases more slowly with $k$ than the other concentration bounds do. This is consistent with its known reliability for large values of $k$. Nevertheless, for the range of values considered for $k$, Agrawal's bounds are our best concentration bounds. The two versions are again very close to each other, both for one and two samples. For this reason, we recommend using $\mathcal M^3_{k,n}$ and $\widetilde{\mathcal M}^{3,\star}_{k,n,m}$, whose expression is simpler than $\mathcal M^2_{k,n}$ and $\widetilde{\mathcal M}^{2,\star}_{k,n,m}$. 

\section{Conclusion}\label{sec:conclusion}

In this paper, we have been interested in the distribution of relative entropy, in the one- and two-sample cases. We have thus presented the asymptotic distribution along with Berry-Esseen bounds. For finite samples, we also have provided concentration bounds. All these approximations of the true distribution are useful for building statistical tests of divergence. A more applied paper on the topic will follow soon. Our theoretical findings rely on recent advances, including results on the multivariate Berry-Esseen inequality and on reverse Pinsker's inequality. These tools do not yet exhibit the same level of theoretical maturity as classical results such as the univariate Berry-Esseen inequality or Pinsker's inequality. Further advances on these two inequalities, for example regarding a non-uniform version of multivariate Berry-Esseen bounds, would directly strengthen our theoretical bounds and enhance the accuracy of divergence tests based on them.

\bibliographystyle{plain}
\bibliography{sanov}

\appendix

\section{A useful lemma for proving Theorem~\ref{thm:TCL}}\label{sec:lem_cubicEquation}

\begin{lem}\label{lem:cubicEquation}
The real solutions of the cubic equation $a x^3+x^2-d= 0$ are 
\begin{equation}\label{eq:Cardan_3racines}
\frac{1}{3a}\left[2\cos\left(\frac{1}{3}\arccos\left(\frac{27a^2d-2}{2}\right)-\frac{2r\pi}{3}\right)-1\right],
\end{equation}
with $r\in\{0,1,2\}$, when $d\leq 4/27a^2$, and
$$\sqrt[3]{ -\frac{1}{27a^3} + \frac{d}{2a} + \sqrt{\frac{-4d+27d^2a^2}{108 a^4}} }
+ \sqrt[3]{ -\frac{1}{27a^3} + \frac{d}{2a} - \sqrt{\frac{-4d+27d^2a^2}{108 a^4}} }
- \frac{1}{3a}$$
when $d> 4/27a^2$. Moreover, when $d>0$ and $a\rightarrow 0$, we get three real roots. One diverges to $+\infty$ if $a<0$ and to $-\infty$ if $a>0$. The other two behave like:
$$\left\{\begin{array}{ccl}
x_0 & = & \sqrt{d}-\frac{d}{2} a+\mathcal O(a^2) \\
x_1 & = & -\sqrt{d}-\frac{d}{2} a+\mathcal O(a^2). 
\end{array}\right.$$
\end{lem}

\begin{proof}
Defining $y$ by $x+1/3a$, we get the following equation in $y$: $y^3+py+q=0$, where $p=-1/3a^2$ and $q=(2/27a^3)-(d/a)$. The discriminant $-(p/3)^3-(q/2)^2$ is equal to $(4d-27d^2a^2)/108 a^4$. There are two cases, depending on the value of $d$. First, if $d\leq 4/27a^2$, the discriminant is non-negative and there are three real roots (given below for the equation in $x$), among which one is double when the discriminant is zero, after Cardano's formula, 
$$\frac{1}{3a}\left[2\cos\left(\frac{1}{3}\arccos\left(\frac{27a^2d-2}{2}\right)-\frac{2r\pi}{3}\right)-1\right],$$
where $r\in\{0,1,2\}$. If $d> 4/27a^2$, the discriminant is non-negative and there is only one real root for the equation in $x$,
$$\sqrt[3]{ -\frac{q}{2} + \sqrt{ \left( \frac{q}{2} \right)^2 + \left( \frac{p}{3} \right)^3 } }
+ \sqrt[3]{ -\frac{q}{2} - \sqrt{ \left( \frac{q}{2} \right)^2 + \left( \frac{p}{3} \right)^3 } }
- \frac{1}{3a},$$
that is, when replacing $p$ and $q$ by their expression in $a$ and $d$,
$$\sqrt[3]{ -\frac{1}{27a^3} + \frac{d}{2a} + \sqrt{\frac{-4d+27d^2a^2}{108 a^4}} }
+ \sqrt[3]{ -\frac{1}{27a^3} + \frac{d}{2a} - \sqrt{\frac{-4d+27d^2a^2}{108 a^4}} }
- \frac{1}{3a}.$$
When $a\rightarrow 0$, the leading term in the discriminant is $d/27a^4$, which is positive if $d>0$. Therefore, the solutions are like in equation~\eqref{eq:Cardan_3racines}. We easily get the limit 
$$\underset{a\rightarrow 0}{\lim}\ 2\cos\left(\frac{1}{3}\arccos\left(\frac{27a^2d-2}{2}\right)-\frac{2r\pi}{3}\right)-1=\left\{\begin{array}{ll}
0 & \text{if } r\in\{0,1\} \\
-3 & \text{if } r=2.
\end{array}\right.$$
In this asymptotic case, the root with $r=2$ will diverge to $+\infty$ if $a<0$ and to $-\infty$ if $a>0$. For the two other roots, we apply a perturbative expansion. We start with the simplified problem corresponding to $a=0$, that is $x^2-d=0$. The two solutions are $\sqrt{d}$ and $-\sqrt{d}$. Next, we consider the perturbation of the two roots: 
$$\left\{\begin{array}{ccl}
x_0 & = & \sqrt{d}+\beta_0 a+\mathcal O(a^2) \\
x_1 & = & -\sqrt{d}+\beta_1 a+\mathcal O(a^2). 
\end{array}\right.$$
Plugging $x_0$ in the cubic equation, we get
$$ad^{3/2}+2d^{1/2}\beta_0 a=\mathcal O(a^2),$$
so that $\beta_0=-d/2$. Using the same approach for $x_1$, we find as well that $\beta_1=-d/2$. We therefore get the result displayed in the lemma.
\end{proof}

\section{Proof of Theorem~\ref{thm:TCL_two-sample} and Proposition~\ref{pro:BerryEsseen_two-sample}}\label{sec:appendix_thm_TCL_two-sample}

We start by the proof of Theorem~\ref{thm:TCL_two-sample}.

\begin{proof}
As seen in the proof of Theorem~\ref{thm:TCL}, the vector whose $i$-th component, for $i\in\llbracket 1,k\rrbracket$, is $\sqrt{n}(\widehat p_{n,i}-p_i)/\sqrt{p_i}$ converges toward a Gaussian vector $\mathbf G_1$ of covariance matrix $\bm{\Gamma}_k$. We have the same result when considering $\sqrt{m}(\widehat q_{m,i}-p_i)/\sqrt{p_i}$,  with the convergence toward the Gaussian vector $\mathbf G_2$ of covariance matrix $\bm{\Gamma}_k$ and independent of $\mathbf G_1$. Therefore, we get, for $n,m\rightarrow\infty$,
$$\begin{array}{ccl}
\sqrt{\frac{nm}{n+m}}\left(\begin{array}{c}
\frac{\widehat p_{n,1}-\widehat q_{m,1}}{\sqrt{p_1}} \\
\vdots \\
\frac{\widehat p_{n,k}-\widehat q_{m,k}}{\sqrt{p_k}}\end{array}\right) 
 & = &
\sqrt{n}\sqrt{\frac{m}{n+m}}\left(\begin{array}{c}
\frac{\widehat p_{n,1}-p_1}{\sqrt{p_1}} \\
\vdots \\
\frac{\widehat p_{n,k}-p_k}{\sqrt{p_k}}\end{array}\right)
-
\sqrt{m}\sqrt{\frac{n}{n+m}}\left(\begin{array}{c}
\frac{\widehat q_{m,1}-p_1}{\sqrt{p_1}} \\
\vdots \\
\frac{\widehat q_{m,k}-p_k}{\sqrt{p_k}}\end{array}\right) \\
 & 
\overset{\text{d}}{\longrightarrow} & \sqrt{1-\lambda}\mathbf G_1-\sqrt{\lambda}\mathbf G_2,
\end{array}$$
which, by independence, is a Gaussian vector of covariance matrix $\bm{\Gamma}_k$.

The successive derivatives of the relative entropy are $\partial_{p_i} D_{KL}(\mathbf p\Vert \mathbf q)=1+\log(p_i/q_i)$, $\partial_{q_i} D_{KL}(\mathbf p\Vert \mathbf q)=-p_i/q_i$, $\partial^2_{p_ip_i} D_{KL}(\mathbf p\Vert \mathbf q)= 1/p_i$, $\partial^2_{q_iq_i} D_{KL}(\mathbf p\Vert \mathbf q)= p_i/q_i^2$, and $\partial^2_{p_iq_i} D_{KL}(\mathbf p\Vert \mathbf q)=-1/q_i$. Therefore, noting $\delta_{n,i}=\widehat p_{n,i}-p_i$ and $\gamma_{m,i}=\widehat q_{m,i}-p_i$, a second-order Taylor expansion gives
$$\begin{array}{ccl}
D_{KL}(\widehat{\mathbf p}_{n}\Vert \widehat{\mathbf q}_{m}) & = & \sum_{i=1}^k \left[\delta_{n,i}-\gamma_{m,i}+\frac{\delta_{n,i}^2}{2p_i}+\frac{\gamma_{m,i}^2}{2p_i}-\frac{\delta_{n,i}\gamma_{m,i}}{p_i} + o\left(\delta_{n,i}^2+\gamma_{m,i}^2\right)\right] \\
 & = & \sum_{i=1}^k \frac{(\delta_{n,i}-\gamma_{m,i})^2}{2p_i} + o\left(\delta_{n,i}^2+\gamma_{m,i}^2\right) \\
 & = & \sum_{i=1}^k \frac{(\widehat p_{n,i}-\widehat q_{m,i})^2}{2p_i} + o\left(\delta_{n,i}^2+\gamma_{m,i}^2\right).
\end{array}$$
The leading term of this expansion is proportional to the Euclidean norm of the vector whose limit it the above Gaussian of covariance $\bm{\Gamma}_k$. The framework is so exactly the same as in the proof of Theorem~\ref{thm:TCL} and we can conclude that
$$2\frac{nm}{n+m}D_{KL}(\widehat{\mathbf p}_{n}\Vert \widehat{\mathbf q}_{m}) \overset{\text{d}}{\longrightarrow} \chi^2_{k-1}.$$
\end{proof}

We now prove Proposition~\ref{pro:BerryEsseen_two-sample}.

\begin{proof}
We define, for $j\in\llbracket 1,n+m\rrbracket$, 
$$\mathbf Y_j=\left(\frac{\indic_{X_j\in\Omega_1}-p_1}{\sqrt{(n+m)p_1}},...,\frac{\indic_{X_j\in\Omega_k}-p_k}{\sqrt{(n+m)p_k}}\right)^t$$
and
$$\widetilde{\mathbf Y}_j=\left\{\begin{array}{ll}
\sqrt{\frac{m}{n}}\mathbf Y_j & \text{if } j\in\llbracket 1,n\rrbracket \\
-\sqrt{\frac{n}{m}}\mathbf Y_j & \text{if } j\in\llbracket n+1,n+m\rrbracket.
\end{array}\right.$$
Then, $\widetilde{\mathbf Y}_1,...,\widetilde{\mathbf Y}_{n+m}$ are independent of each other and such that, noting $\widetilde{\mathbf W}=\sum_{j=1}^{n+m}\widetilde{\mathbf Y}_j$, we have $\E(\widetilde{\mathbf Y}_j)=0$ and, after Theorem~\ref{thm:TCL_two-sample}, $\sum_{j=1}^{n+m} \E(\widetilde{\mathbf Y}_j\widetilde{\mathbf Y}_j^t)=\E(\widetilde{\mathbf W}\widetilde{\mathbf W}^t)=\bm{\Gamma}_k$, with the same $\bm{\Gamma}_k=\mathbf V\mathbf V^t$ as in the proof of Theorem~\ref{thm:TCL}. Therefore, we can use Rai\v{c}'s theorem again~\cite[Theorem 1.1]{Raic}, applied to $\widetilde{\mathbf U}=\mathbf V^t\widetilde{\mathbf W}$ and $\widetilde{\mathbf T}_j=\mathbf V^t\widetilde{\mathbf Y}_j$, $\widetilde{\mathbf T}_j$ having the same Euclidean norm as $\widetilde{\mathbf Y}_j$. Noting 
$$\widetilde \Theta=\widetilde{\mathbf W}^t\widetilde{\mathbf W}=\frac{nm}{n+m}\sum_{i=1}^k \frac{\left(\widehat p_i-\widehat q_i\right)^2}{p_i},$$
we can thus write, like in the proof of Theorem~\ref{thm:TCL},
\begin{equation}\label{eq:moment3_normeEucl_twosample}
\left|\proba\left(\widetilde \Theta\leq x\right)-F_{\chi^2_{k-1}}(x)\right| \leq \left(42(k-1)^{1/4}+16\right)\sum_{j=1}^{n+m}\E\left[\left(\widetilde{\mathbf Y}_j^t\widetilde{\mathbf Y}_j\right)^{3/2}\right].
\end{equation}
We know, from equation~\eqref{eq:moment3_normeEucl} the expression of $\E\left[\left(\mathbf Y_j^t\mathbf Y_j\right)^{3/2}\right]$, from which we deduce that 
$$\E\left[\left(\widetilde{\mathbf Y}_j^t\widetilde{\mathbf Y}_j\right)^{3/2}\right]=\left\{\begin{array}{ll}
\frac{m^{3/2}}{n^{3/2}(n+m)^{3/2}}\sum_{u=1}^k \frac{(1-p_u)^{3/2}}{p_u^{1/2}} & \text{if } j\in\llbracket 1,n\rrbracket \\
\frac{n^{3/2}}{m^{3/2}(n+m)^{3/2}}\sum_{u=1}^k \frac{(1-p_u)^{3/2}}{p_u^{1/2}} & \text{if } j\in\llbracket n+1,n+m\rrbracket.
\end{array}\right.$$
Therefore 
$$\begin{array}{ccl}
\sum_{j=1}^{n+m}\E\left[\left(\widetilde{\mathbf Y}_j^t\widetilde{\mathbf Y}_j\right)^{3/2}\right] & = & \left(\frac{m^{3/2}}{n^{1/2}(n+m)^{3/2}} + \frac{n^{3/2}}{m^{1/2}(n+m)^{3/2}}\right)\sum_{u=1}^k \frac{(1-p_u)^{3/2}}{p_u^{1/2}} \\
 & = & \frac{m^2+n^2}{(nm)^{1/2}(n+m)^{3/2}}\sum_{u=1}^k \frac{(1-p_u)^{3/2}}{p_u^{1/2}}
\end{array}$$
and, combining it with equation~\eqref{eq:moment3_normeEucl_twosample}, we finally obtain the result displayed in Proposition~\ref{pro:BerryEsseen_two-sample}.
\end{proof}

\section{Proof of Proposition~\ref{pro:Agrawal}}\label{sec:appendix_pro_Agrawal}

\begin{proof}
We know that the moment-generating function of $D_{\text{KL}}\left(\widehat{\mathbf p}_n\Vert \mathbf p\right)$ is upper bounded by 
$$M(t)=\left(\sum_{j=0}^n \frac{n!}{n^{2j}(n-j)!}t^j\right)^{k-1},$$
for $t\in[0,n)$~\cite[Proposition II.2, Lemmas II.4 and II.5]{Agrawal}. So, using Chernoff inequality, we get the first bound  displayed in Proposition~\ref{pro:Agrawal}, $\mathcal M^{1}_{k,n}(x)$. The last bound, $\mathcal M^{3}_{k,n}(x)$, is Agrawal's bound, which relies on the fact that $M(t)\leq M_{\text{Agrawal}}(t)=\left(1-t/n\right)^{-k+1}$ and on the optimisation in $t$ of the Chernoff bound $\inf_{t\in[0,n)}e^{-tx}M_{\text{Agrawal}}(t)$, the minimum being reached for $t$ equal to $t^{\star}=n-(k-1)/x\in[0,n)$~\cite[Theorem I.2]{Agrawal}. Using this same value $t^{\star}$ in $M(t)$, we get the middle bound, $\mathcal M^{2}_{k,n}(x)$, which is higher than $\mathcal M^{1}_{k,n}(x)$ because $t^{\star}\in[0,n)$, and lower than $\mathcal M^{3}_{k,n}(x)$ because $M(t)\leq M_{\text{Agrawal}}(t)$ for all $t\in[0,n)$, including $t^{\star}$.
\end{proof}

\section{Proofs for two-sample concentration inequalities}

\subsection{Proof of Proposition~\ref{pro:inegKL}}\label{sec:appendix_inegKL}

\begin{proof}
Noting $\Vert \mathbf x\Vert_1=\sum_{i=1}^k |x_i|$, we have, by triangular inequality $\Vert \mathbf x-\mathbf y\Vert_1\leq \Vert \mathbf x\Vert_1+\Vert \mathbf y\Vert_1$, so, combining this with Young's inequality, we get
\begin{equation}\label{eq:inegtriangnorme1}
\Vert \mathbf x-\mathbf y\Vert_1^2\leq 2\Vert \mathbf x\Vert_1^2+2\Vert \mathbf y\Vert_1^2.
\end{equation}
Using successively the right part of formula~\eqref{eq:Pinsker}, formula~\eqref{eq:inegtriangnorme1}, and the left part of formula~\eqref{eq:Pinsker}, we get
$$\begin{array}{ccl}
D_{\text{KL}}\left(\mathbf p\Vert \mathbf q\right) & \leq & \left(\frac{1}{2\min_i q_i}-\min_i \frac{p_i}{2q_i}\right) \Vert \mathbf p-\mathbf q\Vert_1^2 \\
 & \leq & \left(\frac{1}{\min_i q_i}-\min_i \frac{p_i}{q_i}\right) \left(\Vert \mathbf p-\mathbf r\Vert_1^2+\Vert \mathbf q-\mathbf r\Vert_1^2\right) \\
 & \leq & \left(\frac{2}{\min_i q_i}-2\min_i \frac{p_i}{q_i}\right) \left(D_{\text{KL}}\left(\mathbf p\Vert \mathbf r\right)+D_{\text{KL}}\left(\mathbf q\Vert \mathbf r\right)\right).
\end{array}$$
\end{proof}

\subsection{Proof of Theorem~\ref{thm:Agrawal_twosample}}\label{sec:appendix_Agrawal_twosample}

\begin{proof}
We note 
$$\beta_{k,m,n} = \frac{2}{\min_{i\in\llbracket 1,k\rrbracket} \overline q_{m,i}}-2\min_{i\in\llbracket 1,k\rrbracket} \frac{\widehat p_{n,i}}{\overline q_{m,i}}\leq\beta_{k,m}.$$
Using Proposition~\ref{pro:inegKL} along with the independence of $\widehat{\mathbf p}_n$ and $\widehat{\mathbf q}_m$, the moment-generating function of the two-sample relative entropy follows, for $t\geq 0$:
$$\begin{array}{ccl}
\E\left[\exp\left(tD_{\text{KL}}\left(\widehat{\mathbf p}_n\Vert \overline{\mathbf q}_m\right)\right)\right]  & \leq &  \E\left[\exp\left(t\beta_{k,m,n} D_{\text{KL}}\left(\widehat{\mathbf p}_n\Vert \mathbf p\right) + t\beta_{k,m,n} D_{\text{KL}}\left(\overline{\mathbf q}_m\Vert \mathbf p\right)\right)\right] \\
 & \leq & \E\left[\exp\left(t\beta_{k,m} D_{\text{KL}}\left(\widehat{\mathbf p}_n\Vert \mathbf p\right)\right)\right] \E\left[\exp\left(t\beta_{k,m} D_{\text{KL}}\left(\overline{\mathbf q}_m\Vert \mathbf p\right)\right)\right],
\end{array}$$
where, by convexity, 
$$D_{\text{KL}}\left(\overline{\mathbf q}_m\Vert \mathbf p\right) \leq \frac{m}{m'}D_{\text{KL}}\left(\widehat{\mathbf q}_m\Vert \mathbf p\right) + \frac{k}{m'}D_{\text{KL}}\left(\mathbf u\Vert \mathbf p\right),$$
with $\mathbf u$ the uniform probability vector and $m'=m+k$.
According to Agrawal's inequality, we thus have, for $t\in[0,\min(m',n)/\beta_{k,m})$ \cite[Theorem 1.3]{Agrawal}:
\begin{equation}\label{eq:preuve2echconc_0}
\E\left[\exp\left(tD_{\text{KL}}\left(\widehat{\mathbf p}_n\Vert \overline{\mathbf q}_m\right)\right)\right] \leq \exp\left(t\beta_{k,m}\zeta_{k,m}\right)\left(\frac{1}{(1-t\beta_{k,m}/m')(1-t\beta_{k,m}/n)}\right)^{k-1}.
\end{equation}
Therefore, by a substitution $s=t\beta_{k,m}$ and Chernoff inequality, we obtain, for $x>0$, $\proba\left(D_{\text{KL}}\left(\widehat{\mathbf p}_n\Vert \widehat{\mathbf q}_n\right)\geq x\right) \leq \inf_{s\in[0,\min(m',n))} \exp(f(s))$, with
$$f(s)= s\left(\zeta_{k,m}-\frac{x}{\beta_{k,m}}\right)-(k-1)\log\left(1-\frac{s}{m'}\right)-(k-1)\log\left(1-\frac{s}{n}\right).$$

We have $f(0)=0$ and $\lim_{s\rightarrow \min(m',n)}f(s)=+\infty$. The derivative of $f$ is
$$f'(s)=\zeta_{k,m}-\frac{x}{\beta_{k,m}}-(k-1)\left(\frac{1}{s-m'}+\frac{1}{s-n}\right).$$
Noting $\lambda=-(\zeta_{k,m}-x/\beta_{k,m})/(k-1)$, $s$ is a zero of $f'$ iff
$$\lambda s^2+(2-\lambda(n+m')) s + \lambda m' n-m'-n =0,$$
equation whose discriminant is $4+\lambda^2(m'-n)^2>0$, leading to the two possible roots 
$$\left\{\begin{array}{lll}
s^{\star} & = & \frac{\lambda(n+m')-2-\sqrt{4+\lambda^2(m'-n)^2}}{2\lambda} \\
s^{\bullet} & = & \frac{\lambda(n+m')-2+\sqrt{4+\lambda^2(m'-n)^2}}{2\lambda}. 
\end{array}\right.$$
By symmetry, one can assume $m'\leq n$. The condition $s^{\star}\in[0,\min(m',n))$ gives, for the upper bound,
$$\lambda (n+m')-2-\sqrt{4+\lambda^2(m'-n)^2} < 2\lambda m',$$
that is
\begin{equation}\label{eq:preuve2echconc_1}
\lambda (n-m')-2<\sqrt{4+\lambda^2(m'-n)^2}.
\end{equation}
When $\lambda(n-m')< 2$, formula~\eqref{eq:preuve2echconc_1} always holds. If $\lambda(n-m') \geq 2$, then, after considering the square, formula~\eqref{eq:preuve2echconc_1} simplifies to $-4\lambda(n-m')< 0$, which always holds since $\lambda(n-m') \geq 2$. Now, the lower constraint $s^{\star}\geq 0$ leads to 
$$\lambda (m'+n)-2\geq\sqrt{4+\lambda^2(m'-n)^2},$$
that is $\lambda(m'+n)\geq 2$ and $\lambda\geq(m'+n)/mn$. A direct calculation also shows that $(m'+n)/m'n>2/(m'+n)$. On the other hand, the root $s^{\bullet}$ never falls in the interval $[0,\min(m',n))$. Indeed, still assuming $m'\leq n$, the condition $s^{\bullet}<m'$ requires
\begin{equation}\label{eq:preuve2Agrawal_Cond}
\lambda (n-m')-2<-\sqrt{4+\lambda^2(m'-n)^2},
\end{equation}
which is never verified because, by the elementary inequality, $2(4+\lambda^2(m'-n)^2)\geq (2+\lambda|m'-n|)^2$, so that condition~\eqref{eq:preuve2Agrawal_Cond} writes
$$\lambda (n-m')-2<-\frac{1}{\sqrt{2}}(2+\lambda|m'-n|)$$
and $\lambda(n-m')<\sqrt{2}(2-\sqrt{2})/(1+\sqrt{2})<0$, which contradicts the assumptions. 

Therefore, when $\lambda\geq\max(2/(m'+n),(m'+n)/m'n)=(m'+n)/m'n$, then $f'(0)\leq 0$ and the minimum of $f$ in the interval $[0,\min(m',n))$ is reached in $s^{\star}$. If $\lambda<(m'+n)/m'n$, then $f'(0)>0$, with no root of $f'$ in $[0,\min(m',n))$, so that the minimum of $f$ in this interval is reached in 0. This provides us with the bound $\widetilde{\mathcal M}^3_{k,n,m}(x)$, in which $\sigma_{m,n,x}$ is simply equal to $s^{\star}$.

Like in Proposition~\ref{pro:Agrawal}, one can also modify formula~\eqref{eq:preuve2echconc_0} to use the true moment-generating function, so we get $\widetilde{\mathcal M}^1_{k,n,m}(x)$ and $\widetilde{\mathcal M}^2_{k,n,m}(x)$.
\end{proof}

\end{document}